\documentclass[acmtog]{acmart}
\acmSubmissionID{1047}

\usepackage{booktabs} 
\usepackage{bm}
\usepackage{multirow}
\usepackage{subcaption}
\usepackage{bbding}
\usepackage{parskip}
\usepackage{geometry}
\usepackage{xcolor}
\usepackage{adjustbox}
\usepackage{makecell}
\usepackage{array} 
\usepackage{graphicx}  

\usepackage[ruled]{algorithm2e} 

\usepackage{utfsym}
\newcommand{\cmark}{\checkmark}%
\newcommand{\xmark}{\scalebox{0.75}{\usym{2613}}}
\SetAlFnt{\small}
\SetAlCapFnt{\small}
\SetAlCapNameFnt{\small}
\SetAlCapHSkip{0pt}

\acmJournal{TOG}

\acmJournal{TOG}
\acmYear{2025} \acmVolume{44} \acmNumber{4} \acmArticle{} \acmMonth{8} \acmPrice{}\acmDOI{10.1145/3731206}

\acmYear{2025} \acmVolume{44} \acmNumber{4} \acmArticle{} \acmMonth{8} \acmPrice{}\acmDOI{10.1145/3731206}

\newcommand{\method}{Diffuse-CLoC\xspace}
\newcommand{\baselinetable}{
\begin{table*}[t]
\centering
\renewcommand{\arraystretch}{1.1} 
\setlength{\tabcolsep}{6pt} 
\caption{Benchmark for Kinematic-tracking baselines v.s. \method across various tasks.}

\begin{tabular}{lcccccccccc}
\hline
\multirow{2}{*}{Method} &
  \multirow{2}{*}{\begin{tabular}[c]{@{}c@{}}Replan\\ Interval\end{tabular}} &
  \multicolumn{2}{c}{Walk + Perturb} &
  \multicolumn{2}{c}{Forest} &
  Jump &
  \multicolumn{3}{c}{Motion In-Between} \\

  \cline{3-4} \cline{5-6} \cline{7-7} \cline{8-10} 
 &  &
  \% Fall $\downarrow$ &
  FID $\downarrow$ &
  \% Success $\uparrow$ &
  Time $\downarrow$ &
  \% Success $\uparrow$ &
  \% Fall $\downarrow$ &
  MJPE $\downarrow$ &
  MRPE $\downarrow$ \\ \hline
\multirow{5}{*}{Kin+PHC} 
        & 1  & 30 & 0.185 & 69 & 39.72 & 0  & 66 & 0.2 & 0.985 \\
        & 4  & 26 & 0.162 & 38 & 40.31 & 1  & 38 & 0.148 & 0.544 \\
        & 8  & 28 & 0.088 & 83 & 22.57 & 21 & 55 & 0.158 & 0.53 \\
        & 16 & 30 & 0.073 & 68 & 16.87 & 22 & 55 & 0.140 & 0.388 \\
        & 32 & 44 & \textbf{0.038} & 22 & 22.28 & 15 & 60 & 0.191 & 0.939 \\ \hline
\method (ours) & 1  & \textbf{16} & 0.074 & \textbf{96} & \textbf{13.23} & \textbf{71} & \textbf{31} & \textbf{0.116} & \textbf{0.322} \\ \hline
\end{tabular}
\label{tab:baseline}
\end{table*}
}

\newcommand{\combinedtable}{
\begin{table}[tbp]
    \centering
    \renewcommand{\arraystretch}{1.1} 
    \setlength{\tabcolsep}{2pt}
    \caption{Ablation study on horizon lengths, diffusion steps, attention styles, and rolling schemes. * marks the final model choice.}
    \begin{tabular}{cc c cc c}
        \hline
        \multicolumn{2}{c}{Horizon} & Speed & \multicolumn{2}{c}{Forest} & Jump \\ \cline{4-6}
        State & Action & (Hz) & \% Success $\uparrow$ & Task Time $\downarrow$ & \% Success $\uparrow$ \\
        \hline
        \multirow{4}{*}{16} & 1 & 85.47 & 4 & 50.24 & 0 \\
        & 4 & 85.47 & 5 & 34.28 & 0 \\
        & 8 & 85.47 & 1 & 59.87 & 1 \\
        & 16 & 85.47 & 0 & NA & 0 \\
        \hline
        \multirow{4}{*}{$32^*$} & 1 & 77.41 & 57 & 21.83 & 18 \\
        & 4 & 77.41 & 84 & 14.30 & $\textbf{77}$ \\
        & 8 & 77.41 & 80 & $\textbf{13.14}$ & 75 \\
        & $16^*$ & 77.41 & $\textbf{96}$ & 13.23 & 71 \\
        \hline
        \multirow{4}{*}{64} & 1 & 60.06 & 20 & 34.67 & 0 \\
        & 4 & 60.06 & 56 & 23.77 & 5 \\
        & 8 & 60.06 & 18 & 15.31 & 33 \\
        & 16 & 60.06 & 19 & 27.84 & 6 \\
        \hline
        \hline
        \multicolumn{6}{l}{\small{Rolling Enabled}} \\ 
        \hline
        $\cmark^*$ & $\cmark^*$ & 77.41 & $\textbf{96}$ & 13.23 & 71 \\ 
        $\cmark$   & $\xmark$  & 58.06 & 79 & 14.75 & $\textbf{76}$ \\ 
        $\xmark$  & $\cmark$ & 58.06 & $\textbf{96}$ & $\textbf{10.62}$ & 43 \\ 
        $\xmark$  & $\xmark$  & 58.06 & 95 & 10.72 & 40 \\ 
        \hline
        \hline
        \multicolumn{6}{l}{\small{Attention Style}} \\ 
        \hline
        \multicolumn{2}{c}{\textcolor{black}{Full}} & 77.41 & 58 & 15.08 & 53 \\
        \multicolumn{2}{c}{Diffuser} & 77.41 & 0 & NA & 0 \\
        \hline
    \end{tabular}
    \label{tab:combined}
\end{table}
}

\begin{document}
\title{Diffuse-CLoC: Guided Diffusion for Physics-based Character Look-ahead Control}

\author{Xiaoyu Huang}
\authornote{Joint First Authors.}
\orcid{0009-0005-0714-3711}
\affiliation{%
 \institution{University of California, Berkeley}
 \country{USA}}
 \affiliation{%
 \institution{RAI Institute}
 \country{USA}
 }
\email{x.h@berkeley.edu}

\author{Takara Truong}
\authornotemark[1]
\authornote{Corresponding Author.}

\affiliation{%
 \institution{Stanford University}
 \country{USA}
}
 \affiliation{%
 \institution{RAI Institute}
 \country{USA}
 }
\email{takaraet@stanford.edu}
\author{Yunbo Zhang}
\affiliation{%
\institution{RAI Institute}
\country{USA}
}
\email{yzhang@theaiinstitute.com}
\author{Fangzhou Yu}
\affiliation{%
 \institution{RAI Institute}
 \country{USA}
}
\email{fyu@theaiinstitute.com}
\author{Jean Pierre Sleiman}
\affiliation{%
 \institution{RAI Institute}
 \country{USA}
 }
\email{jsleiman@theaiinstitute.com}
\author{Jessica Hodgins}
\affiliation{%
 \institution{RAI Institute}
 \country{USA}
}
\email{jkh@cs.cmu.edu}
\author{Koushil Sreenath}
\affiliation{%
 \institution{University of California, Berkeley}
 \country{USA}}
\affiliation{%
 \institution{RAI Institute}
 \country{USA}
}
\email{ksreenath@theaiinstitute.com}
\author{Farbod Farshidian}
\affiliation{%
 \institution{RAI Institute}
 \country{USA}
}
\email{ffarshidian@theaiinstitute.com}

\begin{abstract}


We present Diffuse-CLoC, a guided diffusion framework for physics-based look-ahead control that  enables intuitive, steerable, and physically realistic motion generation. While existing kinematics motion generation with diffusion models offer intuitive steering capabilities with inference-time conditioning, they often fail to produce physically viable motions. In contrast, recent diffusion-based control policies have shown promise in generating physically realizable motion sequences, but the lack of kinematics prediction limits their steerability. Diffuse-CLoC addresses these challenges through a key insight:  modeling the joint distribution of states and actions within a single diffusion model makes action generation steerable by conditioning it on the predicted states. This approach allows us to leverage established conditioning techniques from kinematic motion generation while producing physically realistic motions. As a result, we achieve planning capabilities without the need for a high-level planner. Our method handles a diverse set of unseen long-horizon downstream tasks through a single pre-trained model, including static and dynamic obstacle avoidance, motion in-betweening, and task-space control. Experimental results show that our method significantly outperforms the traditional hierarchical framework of high-level motion diffusion and low-level tracking.

\end{abstract}

%
%




%
%

\begin{teaserfigure}
  \includegraphics[width=\textwidth]{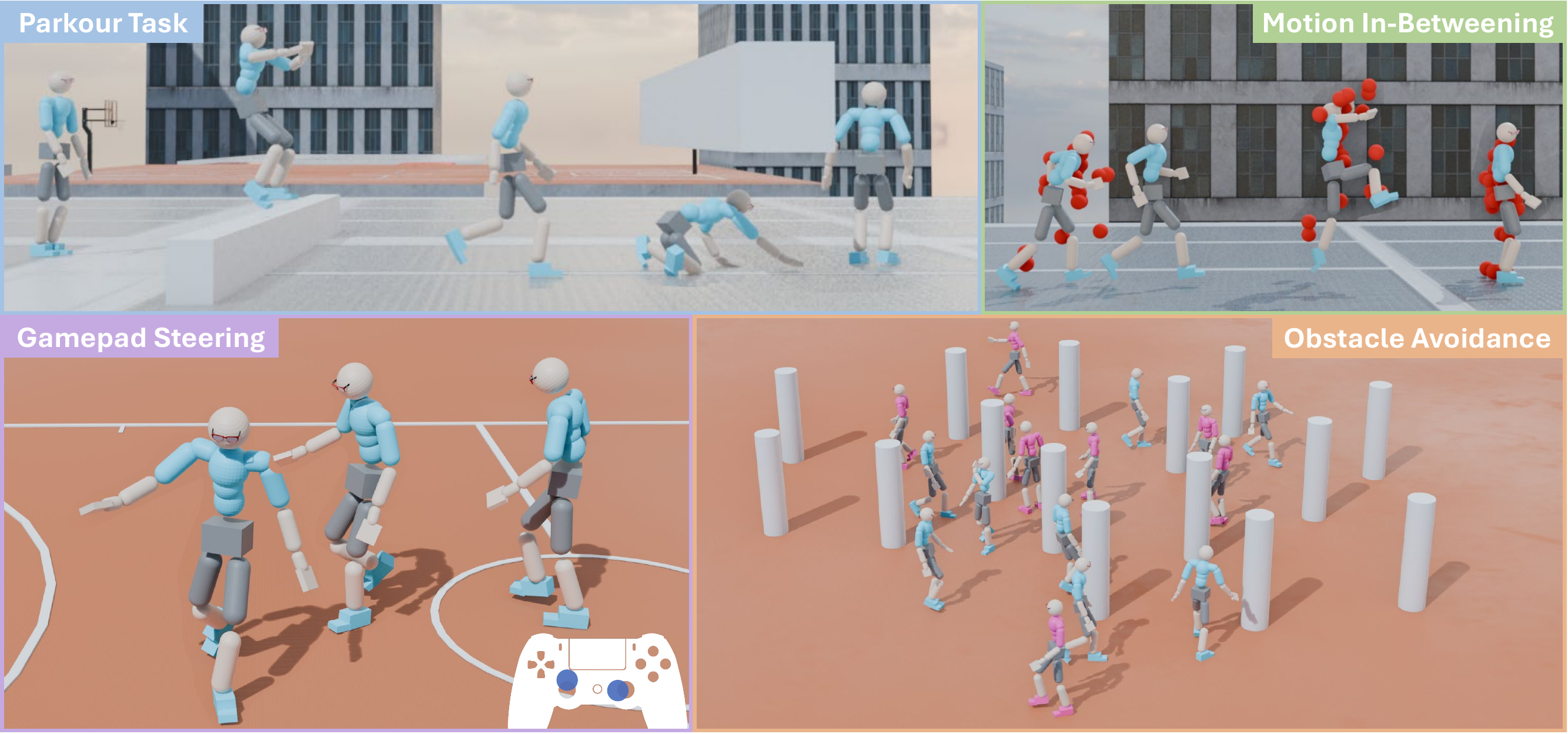}
  \centering
  \caption{\method, demonstrates impressive versatility across a wide range of unseen physics-based tasks using the same pre-trained diffusion model without the need for retraining or finetuning. These tasks include static and dynamic obstacle navigation with 16 characters, agile motion obstacle avoidance, gamepad control guided by a classifier, and physics-based motion inbetweening through inpainting.}
  \label{fig:teaser}
\end{teaserfigure}

\maketitle

\section{Introduction}
\label{sec:intro}

The goal of physics-based character animation has long been the synthesis of dynamically realistic motions that can both react to its environment and respond to precise user control. Such capabilities have broad applications across gaming, virtual/augmented reality, and robotics. 

Recent advances in kinematic motion diffusion models have shown promise in motion synthesis through conditioning on partial states~\cite{yi_AMDM}, text~\cite{guy_MDM}, and music \cite{jonathan_EDGE}. Furthermore, the same kinematics motion diffusion model can be reused on unseen downstream tasks via inference-time conditioning such as \emph{classifier-guidance} or \emph{inpainting} \cite{Guided-MDM, cohan_flexible_motion_inbetween}. Yet, these approaches fall short of achieving physical realism. To improve physical consistency, a tracking policy can be used to follow the kinematics motions, but ensuring robustness to domain shifts introduced by the kinematics generation remains a challenge. On the other hand, prior works \cite{cheng_diffusionPolicy, takara_PDP, xiayou_Diffuseloco} have focused on predicting actions, ensuring physical realism without requiring extra components.
However, unlike kinematic-based diffusion models, these approaches cannot adapt to unseen downstream tasks without retraining, as inference-time conditioning techniques are infeasible due to the lack of direct comparability between goals defined in state space and synthesized action trajectories.

To bridge the gap between kinematic motion diffusion and action diffusion models, we propose \method, a diffusion model designed for flexible guidance in physics-based look-ahead control. Our goal is to train an end-to-end state-action diffusion model capable of addressing a wide range of unseen, long-horizon downstream tasks for a physics-based character, without the need for retraining.


Our primary insight lies in modeling the joint distribution of states and actions within a single diffusion model, which enables action generation to be conditioned on the predicted states. 
Building on this insight, we develop a transformer-based diffusion architecture and an attention mechanism that includes non-causal attention for states and causal attention for actions. 
We further introduce a rolling scheme inspired by \cite{zhang_Tedi} to improve consistency and speed in auto-regressive policy execution. Ultimately, \method enables the application of established techniques from kinematic motion generation, such as \emph{classifier guidance}, to perform multiple unseen downstream tasks with physics-based characters, including waypoint navigation, dynamic obstacle avoidance such as other characters, and dynamic maneuvers like jumping over obstacles and crawling underneath them, as well as \emph{inpainting} for motion in-betweening task (Fig. \ref{fig:teaser}). Our primary contributions include:
\begin{itemize} 
\item A method for training and guiding a diffusion-based policy that enables the completion of multiple unseen downstream tasks in physics-based character control without requiring task-specific fine-tuning or a high-level planner.


\item A novel architecture and attention setup to model the joint distribution of states and actions, allowing conditional action generation via steerable motion synthesis.  


\item An optimized rolling inference scheme for auto-regressive action generation, enabling interactive execution of agile motions, significantly improving consistency and speed. 

\end{itemize}

\section{Related Work}
\begin{figure*}[t]
    \centering
    \includegraphics[width=0.9\linewidth]{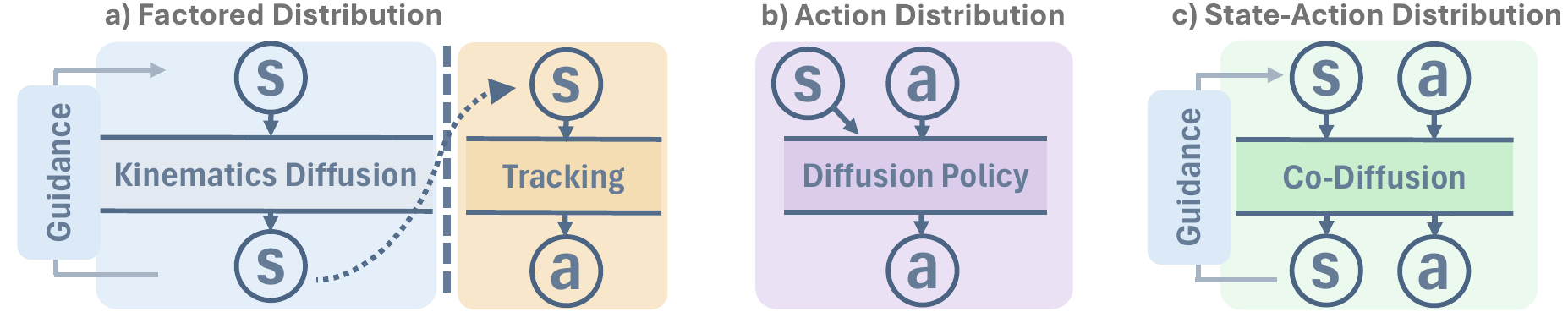}
    \caption{\label{fig:related_works_figure} \textbf{Three Formulations in Physics-based Control using Diffusion}. (a) The factored distribution approach separately learns planning $p(s)$ and control $p(a|s)$, using kinematics planners \cite{Guided-MDM, guy_MDM, zhang_Tedi} and tracking policies \cite{PHC}. Examples of this category are \cite{zhaoming_BoxLoco, guy_closd, ajay_cond_decision_making}. (b) The model-free approach learns $p(a|s)$ only and does not allow inference-time planning \cite{takara_PDP, cheng_diffusionPolicy, xiayou_Diffuseloco}. (c) The joint distribution approach models $p(s,a)$ directly, enabling state-guided action generation. This includes~\cite{janner_diffuser} and \method, where \method demonstrates its effectiveness in complex character control tasks.}
    
\end{figure*}

This section reviews recent advancements across several domains relevant to our approach, including physics-based character animation, diffusion in physics-based control, steering motion synthesis, and inference consistency and speed. An overview of prior works in diffusion based methods are shown in Fig. \ref{fig:related_works_figure}. 

\subsection{Physics-Based Character Animation} 
Early physics-based character animation focused on tracking single motion trajectories \cite{jason_deepMimic}. Subsequent work expanded this to capture motion distributions, such as AMP \cite{jason_AMP}, though these approaches faced challenges in training stability and were limited to small datasets. As an alternative, latent skill embeddings were introduced, leveraging a VAE trained on large datasets  \cite{Yao2022ControlVAE, yao2023moconvq, jungdom_characterControlVAE, zhu2023neural}, which can also be paired with a high-level controller to complete downstream tasks \cite{jason_ASE, zhu2023neural, jungdom_characterControlVAE}.

Recent motion tracking controllers \cite{PHC, agon_vmp} have achieved notable success, leading to approaches that decompose the problem into two parts: a kinematic motion synthesis model and a universal tracking controller for the synthesized motion \cite{zhaoming_BoxLoco, guy_closd}. However, the tracking controller's performance heavily depends on the quality of the kinematic trajectory, which can include artifacts like floating bodies, foot sliding, and penetration. 

To address this challenge, RobotMDM \cite{agon_robotMDM} fine-tunes a motion diffusion model using the value function of an RL tracker to produce more trackable motions. Similarly, \cite{Gaoge_ReinDiffuse, Zhuo_morph} refine diffusion models directly with PPO-based reinforcement learning guided by an RL tracker’s reward. Other methods instead fine-tune the tracking controller to handle artifacts created by the diffusion model \cite{guy_closd}.

Despite these advancements, trackers are often sensitive to out of distribution motions and require task-specific fine-tuning for applications like object interaction \cite{guy_closd}. Furthermore, we demonstrate that even minor classifier guidance on kinematic trajectories can hurt the robustness of general tracking controllers.

\subsection{Diffusion in Physics-based Control}

Diffusion policies have been successfully applied in character animation \cite{takara_PDP}, manipulation \cite{cheng_diffusionPolicy}, and locomotion \cite{xiayou_Diffuseloco, birodiff} domains. However, its action-only generation makes inference-time conditioning impractical. In contrast, Diffuser \cite{janner_diffuser} diffuses both states and actions, enabling classifier guidance through a reward formulation though exhibits limited robustness \cite{boyuan_diffusionForcing}. Decision Diffuser \cite{ajay_cond_decision_making} improves robustness by similarly diffusing both states and actions, but discards actions during inference, relying instead on a trained inverse dynamics model. In contrast, our method enables both intuitive guidance directly in \textit{state space} and high-quality conditional generation in \textit{action space}.


\subsection{Steering Motion Synthesis}
This section discusses common methods for steering motion synthesis, enabling pretrained models to be re-used for novel downstream tasks.  Each approach provides a unique mechanism to align the generative process with task-specific requirements or constraints.

\emph{Classifier-free guidance}
enhances sample quality and control by interpolating between an unconditional and a conditional model during the generation process, without requiring an explicit classifier~\cite{jonathan_classifier_free_guidance}.
In character animation, this method has been used to synthesize motions with text conditioning \cite{takara_PDP, guy_MDM}. However, the conditioning signals must be specified during training, and the model must be re-trained for any new conditioning signals. 

\emph{Classifier guidance} 
incorporates conditional information after the training phase by utilizing the gradient of a pre-trained classifier \cite{dhariwal_classifierguidance} or cost function \cite{carvalho2024motionplanningdiffusionlearning}. This technique guides the diffusion model's generative process, steering outputs toward the desired class or minimizing the loss. In character animation, it allows a single pre-trained kinematics model to address multiple downstream tasks, such as obstacle avoidance and reaching \cite{Guided-MDM, janner_diffuser}. Despite its success, classifier guidance has not yet been explored for dynamic tasks in underactuated robotic systems, such as physics-based character animation.

\emph{Inpainting}
involves generating or reconstructing missing regions of the original signal by conditioning the diffusion process on the surrounding known context. 
In character animation, this technique is useful for generating trajectories with partially known states, such as root trajectories~\cite{Guided-MDM} or a set of desired joint trajectories~\cite{jonathan_EDGE}.
 
\emph{Reinforcement learning} has been employed as a high level planner to direct the diffusion process for downstream tasks \cite{yi_AMDM}. When future trajectory predictions are absent, a high-level planner is essential to strategize \cite{yi_AMDM}; however, our method can plan and foresee future events, allowing the accomplishment of downstream tasks without the need for a high-level planner. 

\subsection{Inference Consistency and Speed}
A key challenge for generative planners is ensuring that the generated plans do not conflict during rollouts. Frequent replanning with multimodal trajectory distributions can lead to alternating behavior modes, negatively affecting task performance \cite{ralf_dp_with_constraints}. One approach is to execute a fixed action sequence before replanning \cite{cheng_diffusionPolicy}, but this lacks real-time state feedback, reducing robustness. 
Another solution, Action Chunking \cite{tony_actionChunking}, averages predictions over time for consistency but may yield suboptimal results by averaging dissimilar plans. Alternatively, \cite{ralf_dp_with_constraints} selects the closest trajectory to the previous one but incurs overhead from concurrently sampling multiple trajectories.

Inference speedup can be naïvely achieved by minimizing diffusion steps at the cost of diversity, though techniques such as one-step diffusion \cite{zhendong_onestepdiffusionpolicyfast} could mitigate these tradeoffs. Hybrid methods, which pair fast architectures for initial guesses with slower fine-tuning of models \cite{Tianyu_AAMDM}, balance speed and accuracy. 
Rolling or streaming strategies apply noise based on temporal proximity—less for immediate actions and more for distant ones—enabling faster generation of infinite sequences \cite{zhang_Tedi, høeg2024streamingdiffusionpolicyfast}. Building upon this, we find that rolling addresses both motion consistency and speedup by maintaining diversity for future decisions while solidifying immediate ones.

\section{Method}

\label{sec:method}

\begin{figure*}[t]
    \centering
    \includegraphics[width=\textwidth]{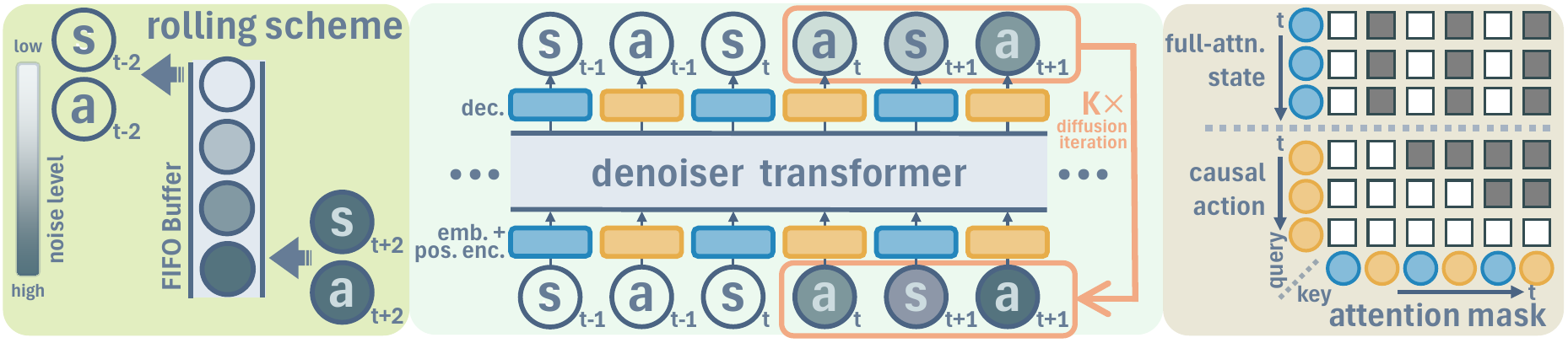}
    \caption{\label{fig:method} \textbf{Framework of \method}. The rolling scheme (Left) is implemented as a FIFO buffer, where at every timestep $t$, a state and action pair of pure noise is pushed into the denoiser, while the earliest clean pair is popped out and used to step the simulator. The denoiser architecture (Middle) denoises the buffer of states and actions with an increasing noise level along the trajectory. The observation $O_t$ is directly inpainted into the sequence, and only noisy future predictions are denoised. Notice that noisy states and actions have different shades, meaning that their noise levels $\bm{k_s}$, $\bm{k_a}$ can be different. The attention mask (Right) shows that state attends to all past and future states but masks out all actions, while action is causal attending to previous states and actions. }
    
\end{figure*}

We propose \method, a novel framework that allows classifier guidance and inpainting in \emph{state space} to generate actions to physically steer the character for various tasks zero-shot. 

Our approach to modeling complex motion behaviors builds upon denoising diffusion probabilistic models (DDPMs)~\cite{ho2020denoising}. At each timestep $t$, our model predicts a trajectory \(\bm{\tau}_{t} = [\bm{a}_{t}, \bm{s}_{t+1}, \bm{a}_{t+1}, \dots, \\ \bm{s}_{t+H}, \bm{a}_{t+H}]\) that contains the current action as well as the $H$ steps of state-action pairs in the future. 
Moreover, we let \( \bm{O}_t \) be the observation history of state-action pairs of length \( N \), as well as the current state, $\bm{s}_t$, i.e., ${\bm{O}_t = [\bm{s}_{t-N}, \bm{a}_{t-N}, \cdots, \bm{s}_{t}]}$. 

We train a prediction network, $\hat{\bm{\tau}}_{t}$ = \( x_{0,\theta}(\bm{\tau}_t^{\bm{k}}, \bm{O}_t, \bm{k}) \) that aims to produce a clean trajectory, where \( \bm{\tau}_t^{\bm{k}} \) is \( \bm{\tau} \) with added Gaussian noise based on the noise level $\bm{k}$. We assign independent noise levels per timestep for states and actions along the trajectory, i.e. $\bm{k} = (\bm{k_s}, \bm{k_a})$, where $\bm{k_s}, \bm{k_a} \in \mathbb{R}^{N+H+1}$. We then follow Stochastic Langevin Dynamics \cite{welling2011bayesian} to sample iteratively,  
\begin{equation}
    \bm{\tau}_{t}^{\bm{k}-1} = \alpha_{\bm{k}} (\bm{\tau}_{t}^{\bm{k}} - \gamma_{\bm{k}} \epsilon_{\theta}(\bm{\tau}_{t}^{\bm{k}}, \bm{O}_{t}, \bm{k}) + \mathcal{N}(0, {\sigma_{\bm{k}}}^2 \bm{I})),
\end{equation}

\noindent
where $\alpha_{\bm{k}}$, $\gamma_{\bm{k}}$, and $\sigma_{\bm{k}}$ are DDPM parameters, and $\epsilon_{\theta}(\cdot) = \frac{1}{\sqrt{1-\alpha_{\bm{k}}}}\bm{\tau}_{t}^{\bm{k}}-\frac{\sqrt{\alpha_{\bm{k}}}}{\sqrt{1-\alpha_{\bm{k}}}}x_{0,\theta}(\cdot)$ is the predicted noise taken away from the noisy trajectory. 
Training of DDPMs is self-supervised, with \( {\bm{k_s}}_i, {\bm{k_a}}_i \sim \mathcal{U}(0, K) \) sampled uniformly up to the maximum diffusion step \( K \). We train with the mean squared error (MSE) loss against the clean trajectory:


\begin{equation}
    \mathcal{L} = MSE( x_{0,\theta}(\bm{\tau}_{t}^{\bm{k}}, \bm{O}_{t}, \bm{k}), \bm{\tau}_t)
\end{equation}

\noindent
`


Diffusion models, though trained unconditionally, can be adapted for conditional generation during inference. By Bayes' rule, the denoiser's learned score function \( \nabla_{\bm{\tau}} \log p(\bm{\tau}) \)~\cite{song2020score} can be extended to the conditional form \(\nabla_{\bm{\tau}} \log p(\bm{\tau} \mid \bm{\tau}^*) = \nabla_{\bm{\tau}} \log p(\bm{\tau}) + \nabla_{\bm{\tau}} \log p(\bm{\tau}^* \mid \bm{\tau})\), where $\bm{\tau}^*$ is the optimal trajectory. Given a cost function \( G_{\bm{\tau}}^c(\bm{\tau}) \), we let the conditional probability of the optimal trajectory to be \( p(\bm{\tau}^* \mid \bm{\tau}) \propto \exp(-G_{\bm{\tau}}^c(\bm{\tau})) \). Then, the posterior gradient, \( \nabla_{\bm{\tau}} \log p(\bm{\tau}^* \mid \bm{\tau}) = -\nabla_{\bm{\tau}} G_{\bm{\tau}}^c(\bm{\tau}) \), naturally aligns with the descending direction of the cost. This approach, formally \emph{classifier guidance}, conditions the diffuser using any cost function \( G_{\bm{\tau}}^c(\bm{\tau}) \) with a computable gradient \( \nabla_{\bm{\tau}} G_{\bm{\tau}}^c(\bm{\tau}) \).

In character animation, classifier guidance is effective in kinematics generation, where cost functions can be readily defined and computed, but not in action space. We address this limitation by a novel co-diffusion of states and actions, allowing kinematic guidance to effectively condition action generation for complex tasks.


\subsection{Co-diffusing States and Actions}
The core idea of our method is co-diffusing state and action, making our diffusion model both an autoregressive motion generator and a control policy. Moreover, we propose a novel transformer architecture, introducing an attention mechanism that allows guidance to propagate from future states to current actions and a loss mask to improve model robustness. This design simultaneously enables long-horizon state predictions and robust actions for complex tasks. 

\emph{Architecture: } Diffuse-CLoC uses a decoder-only transformer architecture, as
shown in Fig.~\ref{fig:method}. In contrast to Diffuser~\cite{janner_diffuser}, where a pair of state and action is combined into a single transformer token, our model takes the state and action input embeddings as separate tokens. 
Both the state and action tokens are concatenated from an MLP embedding and a sinusoidal embedding of their corresponding noise levels $\bm{k_s}$ and $\bm{k_a}$. We use a two-layer MLP for the state encoder and a linear layer for the action encoder. We then add learnable positional encodings to each token. 
The state and action embeddings over time are fed into a GPT-style decoder~\cite{radford2018improving}, followed by layer norm and the corresponding linear layers as output heads. 

\emph{Attention: } 
\method uses a tailored attention mask that differentiates between states and actions. Shown in Fig.~\ref{fig:method} (Right), actions use a causal attention mask, attending only to past states to anchor the trajectory and filter out prediction artifacts in future states, in addition to past actions for regularization. In contrast, states are permitted to attend to future states, enabling future information to backpropagate to current state. Additionally, disabling attention from states to actions simplifies learning without compromising the quality of kinematics prediction.


\emph{Shorter Action Horizon: } 
Predicting long state horizons, such as those extending to about a second, is essential for multi-modality and long-term planning required by certain tasks, but long-term action prediction is challenging due to increased variance~\cite{takara_PDP}. To address this, we limit the action horizon to at most 16 steps and mask long-term future actions in the loss function, focusing on the initial actions only.

\emph{Emphasis Projection: } To enhance state representation, we integrate emphasis projection~\cite{Guided-MDM} to emphasize the global states in the state space. Specifically, we define a projection matrix \( \bm{P} = \bm{A} \bm{B} \), where \(\bm{A}_{ij} \sim \mathcal{N}(0, 1) \), and \( \bm{B} \) a diagonal matrix with entries corresponding to the global states set to \( c > 1 \) and others to one.
Furthermore, unlike~\cite{Guided-MDM}, sensitivity to local states is also preserved by concatenating projected and original states, i.e. $\bm{P} = [\bm{A} \bm{B} \ \ \mathbf{I}]$.

\subsection{Rolling Inference}
During inference, we use receding horizon control and replan at every timestep, executing only the immediate action. 
This approach improves robustness over open-loop execution by allowing the policy to adapt to the latest observations.

However, an issue in constant replanning is that planned trajectories could be inconsistent, causing oscillations. To address this issue, we use a rolling inference scheme. While a similar technique was introduced to generate infinite motions in kinematics space~\cite{zhang_Tedi}, we find it also improves the policy consistency during autoregressive execution.  A FIFO buffer assigns noise levels based on the proximity of trajectory step to the current timestep. At each timestep, new Gaussian noise is added to the buffer while the oldest is removed. This approach leverages previous diffusion results for warmup, ensuring consistency and reducing diffusion steps. Since the the classifier guidance signal diminishes as $\bm{k} \rightarrow 0$, we perform rolling at a higher $\bm{k}$ for stronger guidance, while maintaining consistency and achieving speedup.


\section{Applications}



The goal of pretraining a generative controller is to handle diverse downstream tasks. Unlike hierarchical approaches that separate planners and tracking controllers, co-diffusing states and actions integrates planning and execution during inference through guidance and inpainting. This approach eliminates the need for additional components or training, resulting in a unified policy capable of solving multiple tasks as detailed below.

\subsection{Static Obstacle Avoidance}
\begin{figure*}[t]
    \centering
    \includegraphics[width=0.9 \linewidth]{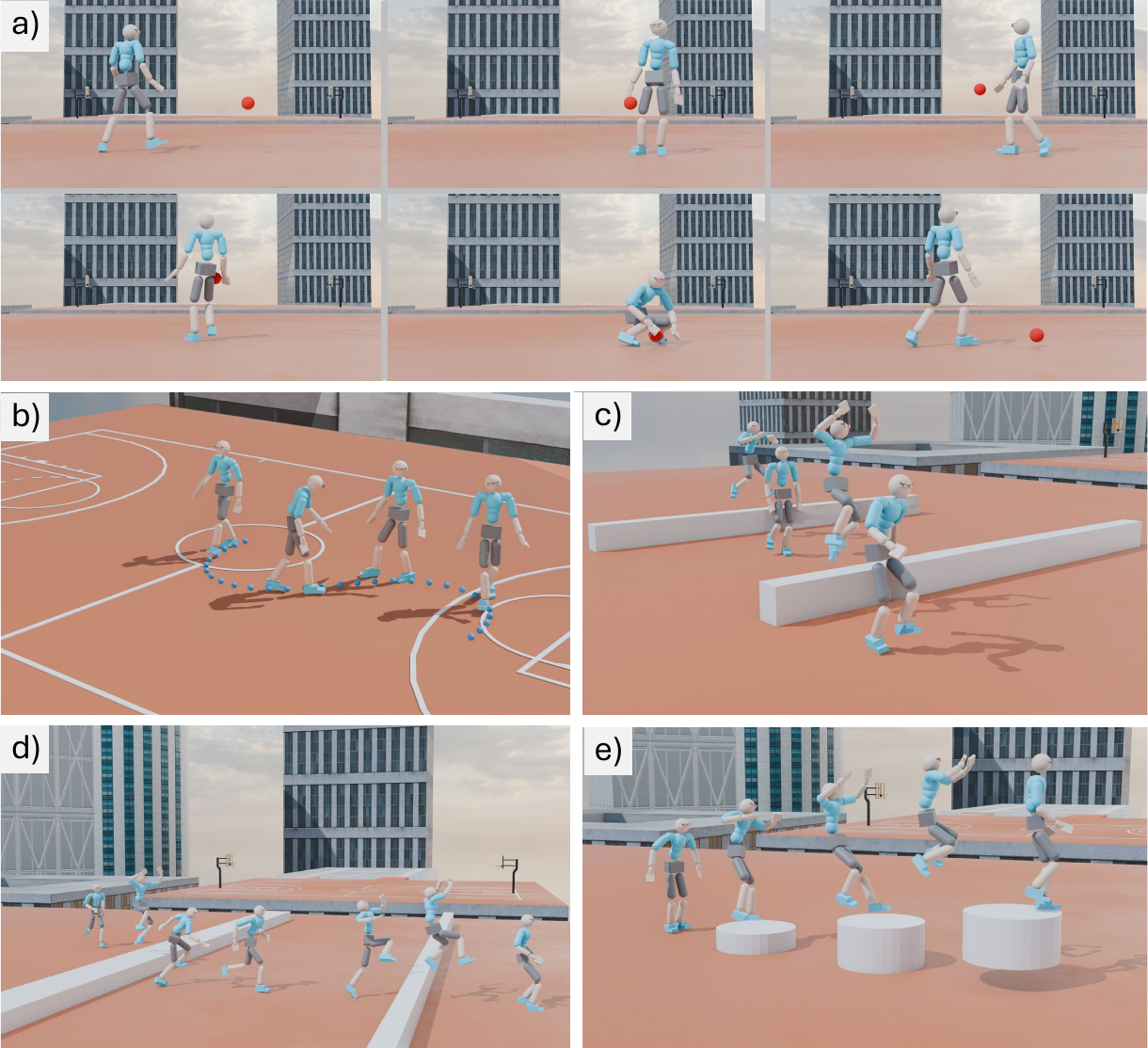}    \caption{\label{fig:downstreamtasks} \textbf{Downstream Tasks} a) Task is to touch the red ball with the character's right hand via classifier guidance. The red ball is re-located when the character achieves the task. b) Root path following via classifier guidance. We can further refine styles by constraining parameters like base velocity and heading. c) Walk and jump over two consecutive obstacles. d) Run and jump over obstacles. e) Jumping on different platforms using penetration cost and waypoint to each platform. f) A sequence of tasks in a single run, including route path following, platform jumping, and reaching a waypoint behind cylindrical barriers. }
    
\end{figure*}

Reacting to obstacles while navigating from A to B can produce the same behavior during inference by introducing a guidance cost that pushes the character away from obstacles:
\begin{equation}
    G_{\bm{\tau}}^{\text{obs}}(\bm{\tau}) = \sum_j \sum_{t' = t}^{t+H} \exp\big(-c \cdot \text{SDF}^{j}(\bm{s}_{t'})\big),
\end{equation}
where $c$ determines the safe distance and $\text{SDF}^j(\cdot)$ is the Signed Distance Function from the $j$-th obstacle. For certain obstacles, such as a block on the ground, interaction (e.g., jumping onto it) may be desired rather than strict avoidance. In this case, we clip the SDF value outside the object to zero, penalizing only when the planned trajectory penetrates the obstacle. To achieve point-goal navigation while avoiding obstacles, the obstacle cost \(G_{\bm{\tau}}^{\text{obs}}(\bm{\tau})\) can be combined with waypoint guidance, $G_{\bm{\tau}}^{\text{wp}}(\bm{\tau}) = \sum_{t' = t}^{t+H}\|P_{\text{root}}(\bm{s}_{t'}) - g\|^2$, where $P_{\text{root}}$ is the mapping from state representation to root position, and $g$ is the waypoint or goal location. We showcase different results using such costs in Fig. \ref{fig:teaser}, \ref{fig:downstreamtasks}c,  \ref{fig:downstreamtasks}d, and \ref{fig:downstreamtasks}e.

\subsection{Dynamic Obstacle Avoidance}
A more advanced obstacle avoidance task involves avoiding not only static obstacles but also each other as dynamic obstacles.
Using the planning horizon, the obstacle 
avoidance cost \(G_{\bm{\tau}}^{\text{obs}}\) can be extended to consider distances for dynamic obstacles at each timestep:  

\begin{equation}
G_{\bm{\tau}, i}^{\text{sa}}(\bm{\tau}) = \sum_{j \neq i} \sum_{t' = t}^{t+H} \exp\big(-c \cdot \|P_{\text{root}}(\bm{s}_{t'}^i) - P_{\text{root}}(\bm{s}_{t'}^j)\|^2\big),
\end{equation}
where \(G_{\bm{\tau}, i}^{\text{sa}}\) is the cost for the \(i\)-th character and \(s_{t'}^j\) is the state of the \(j\)-th character at timestep \(t'\). Fig. \ref{fig:teaser} shows an example of characters navigating through a forest of pillars while avoiding each other. This formulation allows characters to generate smooth trajectories that avoid collisions and pass each other seamlessly.

\subsection{Task Space Control}
Another challenging task is conditioning on target states for specific joints at arbitrary future timesteps within the planning horizon. This problem can be tackled using classifier guidance,
\begin{equation}
G_{\bm{\tau}}^{\text{ts}}(\bm{\tau}) = \sum_{t' \in T} \|P_x(\bm{s}_{t'}) - g_{t'}\|^2,
\end{equation}
where \(T\) is the set of keyframe timesteps, and \(g_{t'}\) represents the target at $t'$. $P_x(\cdot)$ is the mapping from state representation to the task space, such as the root and end-effector positions, velocities, or combinations of them. In this work, we show the following tasks:




\emph{Root Path Following}: Setting the root position as the task space and specifying the target, $g$, as the desired root position along the path ensures precise root path tracking. Fig. \ref{fig:downstreamtasks}b shows a character following an "S" shaped path while switching across styles. 

\emph{Reaching}: Setting one body part as the task space allows the character to generate motions to reach a target $g$ using dataset motions as shown in Fig. \ref{fig:downstreamtasks}a.

\emph{Gamepad Controller Steering}: Shown in Fig. \ref{fig:teaser}, setting the root's velocity, heading, and height as the task space, with $g$ from gamepad inputs, enables real-time character steering.

\subsection{Motion In-Betweening}
In this task, the policy generates dense trajectories between a set of desired keyframes. With sparse keyframes, inpainting often struggles to differentiate clean from noisy states. While previous methods rely on extra masking to label these states~\cite{cohan_flexible_motion_inbetween}, our approach, shown in Fig. \ref{fig:teaser}, can explicitly set the inpainted states to zero noise, naturally differentiating them from other noisy states.

%
\begin{equation}
\bm{s}_t = \hat{\bm{s}}_t, \quad \bm{{k_s}}_t = 0, \quad \forall t \in T,
\end{equation}
where \(\hat{\bm{s}}_t\) represents keyframe states, \(\bm{{k_s}}_t\) is the noise level at timestep \(t\), and \(T\) is the set of keyframe timesteps. 

\baselinetable

\section{Experiments and Results}
\label{sec:experiments}

We evaluate our method in two sections. First, we perform a comparative analysis with conventional hierarchical kinematics-tracking frameworks on various downstream tasks. Second, we ablate on key architectural components to validate our design choices.


\subsection{Data}   
We use a subset of AMASS \cite{amass2019}, comprising 54 motions of walking, running, crawling, and jumping. To track these motions and collect state-action pairs, we employ PHC+~\cite{luo2024universalhumanoidmotionrepresentations} as our motion tracking controller. For each motion, we collect on average 40 rollouts ($\approx 3.5$ hours). Following~\cite{takara_PDP}, we inject action noise during rollout to perturb the state and collect corrective actions. The state and action definitions are as follows:

\paragraph{State}
We use the SMPL skeleton from \cite{PHC}, with $J = 23$ spherical joints. The overall state vector is 165-dimensional, consisting of \emph{Global} and \emph{Local} states defined as the following:

\begin{itemize}
\item \emph{Global States: } root states including position ($\mathbb{R}^{3}$), linear velocity ($\mathbb{R}^{3}$), and rotation ($\mathbb{R}^{3}$) represented as rotation vectors, as in~\cite{zhang2023learning}. These values are expressed relative to the character frame at the \emph{current timestep}.
\item \emph{Local States: } joint states including Cartesian positions ($\mathbb{R}^{3J}$), linear velocities ($\mathbb{R}^{3J}$), and rotations of hands and ankles ($\mathbb{R}^{3\times4}$) also represented as rotation vectors. These quantities are expressed relative to the character frame at \emph{each timestep}.
\end{itemize}

\paragraph{Action}
The action space is a 69-dimensional vector, representing the target joint positions ($\mathbb{R}^{3J}$) for the joint PD controller.

\subsection{Experiment Setup}
We train \method\ using an observation history of $N=4 (\approx 0.13s)$ and predict future states for $H=32 (\approx 1s)$. 
We limit the action horizon to $16$ steps. \textcolor{black}{The model employs a transformer decoder with $6$ layers, $8$ heads, $512$-dimensional embedding, totaling $19.95M$ parameters and requiring approximately 1 GB of GPU memory for inference. Training is conducted with $20$ denoising steps and an attention dropout rate of $p=0.3$}


We optimize using AdamW with a learning rate of \(1 \times 10^{-4}\), weight decay of \(1 \times 10^{-3}\), a 10{,}000-step warmup phase, and a cosine learning rate schedule.
For proper guidance strength, we include state rolling at a noise level of 14, and action rolling at a noise level of 4. We train on a single Nvidia A100 GPU for 1,000 epochs (~24 hours). Inference time is gathered on a Nvidia RTX 4060 GPU with TensorRT acceleration.

\subsubsection{Task \& Metrics}  
We evaluate our method across four tasks:  

\begin{itemize}  
    \item \textbf{Walk-Perturb}: We apply an instantaneous perturbation force sampled between 0 to \(3000 \, \text{N}\) every second during a 30s walk. We assess fall rate and motion quality using FID.  

    \item \textbf{Forest}: We randomly place $15$ cylinder pillars in an \(8 \, \text{m} \times 9 \, \text{m}\) area. The character navigates to a target point on the other side, with success rate and traversal time as metrics.  

    \item \textbf{Jump}: We randomly place a \(0.3\)-\(0.5 \, \text{m}\) tall box obstacle \(1\)-\(2 \, \text{m}\) in front of the character. We measure success rate for clearing the obstacle without falling.

    \item \textbf{Motion In-Betweening}: We generate keyframes by randomly concatenating (without interpolating) three training motions, sampling three keyframes from each at 1-second intervals, and measure fall rate, mean keyframe root error (MRPE), and mean keyframe joint position error (MJPE).  
\end{itemize}  

\noindent For all tasks, a fall is determined if head height drops below \(0.2 \, \text{m}\).  









\subsubsection{Baseline}
The baseline we compare against represents the category of methods that tracks a diffused kinematic trajectory. \textcolor{black}{For the kinematics component, we train a diffusion model similar to the motion generation model in Guided-MDM~\cite{Guided-MDM}, and for tracking we use PHC+ \cite{luo2024universalhumanoidmotionrepresentations} . Since replanning frequency is a key design consideration in hierarchical approaches, we evaluate this baseline (Kin+PHC) across various replanning frequencies.}


\subsubsection{Ablation}
We ablate over critical design choices, including \emph{prediction horizon} for both state and action, as well as the number of diffusion steps. We also compare different attention schemes, including \textcolor{black}{\emph{full attention}, which enables unrestricted attention between all states and actions,} and \emph{diffuser attention}, which restricts attention to each token and its adjacent steps, following \cite{janner_diffuser}.


\begin{figure}[bp]
    \centering
    \includegraphics[width=\linewidth]{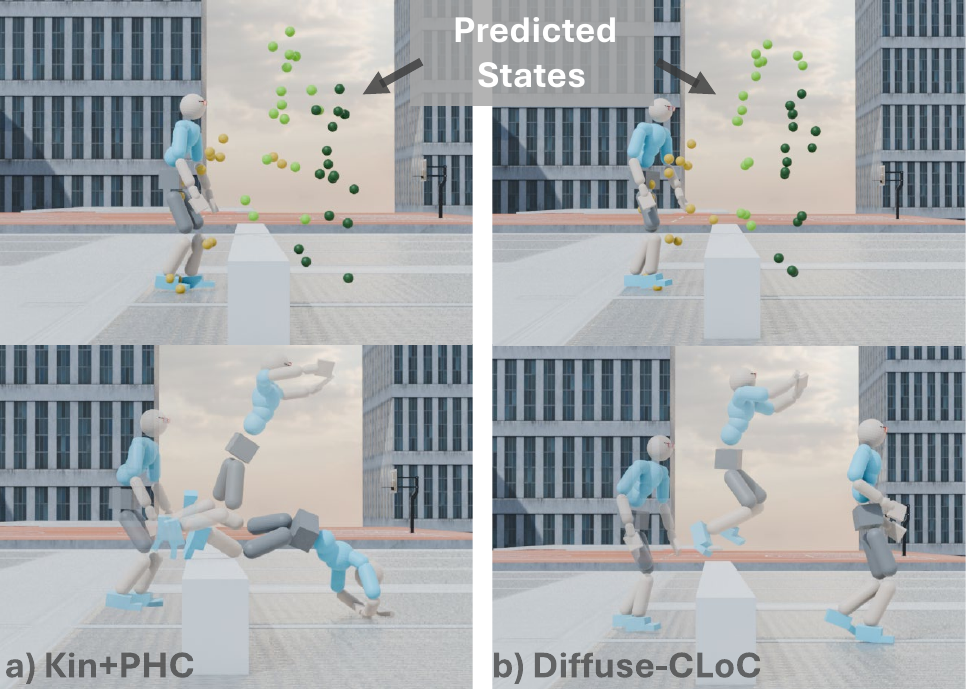}
    \caption{\textbf{Rollouts in \emph{jump} task for Kin+PHC (Left) vs. \method (Right)}. Kin+PHC suffers from artifacts in kinematics prediction and fails in agile motions, while \method remains robust and completes the task. }
    \label{fig:phc_comparison}
\end{figure}

\begin{figure}[bp]
    \centering
        \includegraphics[width=0.8\linewidth]{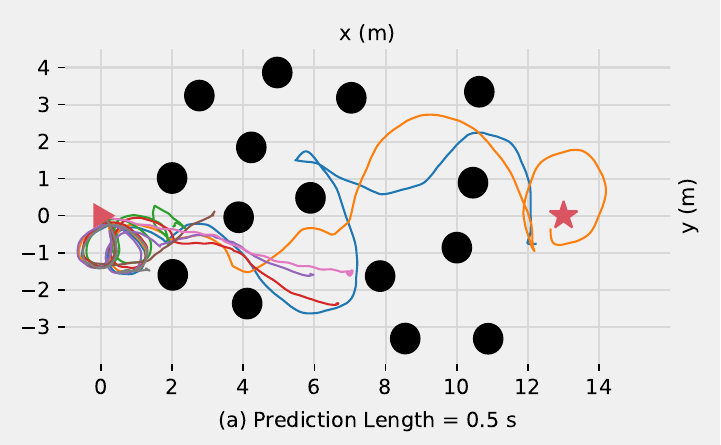}
    \vspace{0.02cm}
        \includegraphics[width=0.8\linewidth]{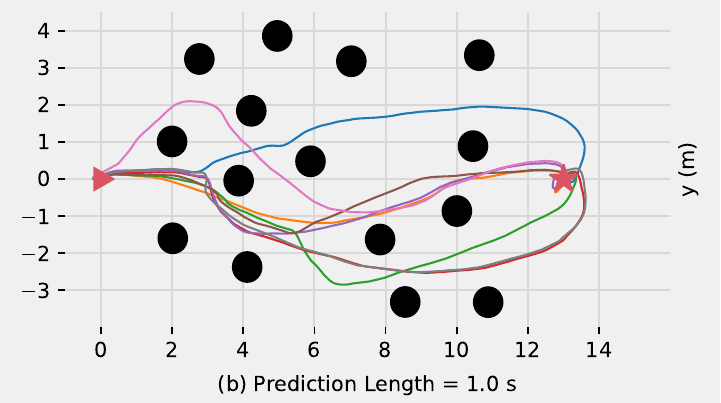}
        \includegraphics[width=0.8\linewidth]{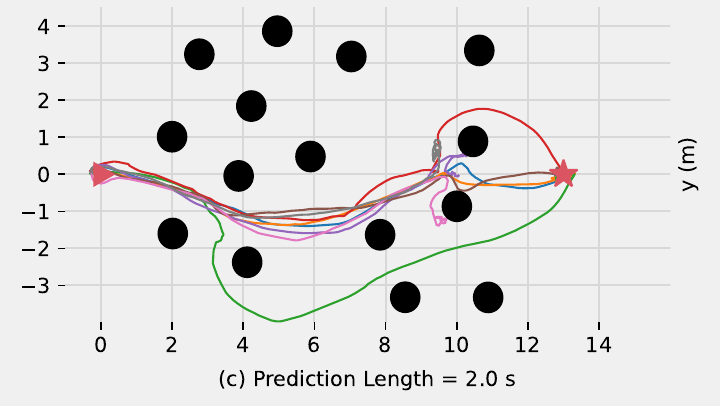}
    \caption{\textcolor{black}{\textbf{Rollouts in the \emph{forest} task for $0.5$s prediction (Top), $1$s prediction (Middle), and $2$s prediction (Bottom).} Starting at the triangle, the character aims to reach the goal marked by a star. The $0.5$s policy circles within the forest and fails to reach the waypoint, while the $1$s policy successfully navigates through the forest with diverse trajectories. In contrast, the $2$s policy sometimes overcommits to suboptimal future paths and collides with obstacles near the goal.}}
    \label{fig:horizon_trajectories}
\end{figure}

\subsection{Comparing Kinematics + Tracking}

We compare Kin+PHC with \method across four tasks, finding that \method consistently outperforms all baseline variants (Table~\ref{tab:baseline}). Kin+PHC exhibits a trade-off between motion quality and task success: slower replan frequencies improve motion smoothness (e.g., FID improves from \(0.185\) at \(1\)-step replan to \(0.038\) at \(32\)-step replan in the \emph{walk+perturb} task) but fails to adapt to abrupt state changes. This results in a higher fall rate, which increases from \(30\%\) at \(1\)-step to \(44\%\) at \(32\)-step intervals. In contrast, \method maintains low FID (\(0.074\)) and reduces the fall rate to \(16\%\), demonstrating robust performance with plausible motion quality.

Kin+PHC's performance also depends heavily on the kinematic planner aligning with the RL tracker's learned distribution, which is vulnerable to inference-time conditioning. In the \emph{jump} task, Kin+PHC achieves only \(22\%\) success at \(16\)-step intervals, as obstacle avoidance guidance generates trajectories outside the tracker's capabilities. Similarly, in \emph{motion in-betweening}, Kin+PHC exhibits higher fall rates (\(66\%\) at \(1\)-step) and less precise motion (MJPE \(0.2\), MRPE \(0.985\)). In contrast, \method achieves \(71\%\) success in \emph{jump} and reduces fall rates to \(31\%\) with lower MJPE (\(0.116\)) and MRPE (\(0.322\)). 


Unlike Kin+PHC, \method uses the joint distribution of states and actions, allowing it to fit over the entire horizon to generate dynamically feasible actions, even with partially corrupted kinematic inputs. By avoiding strict state tracking, \method reduces the infeasible motions from classifier guidance, as shown in Fig.~\ref{fig:phc_comparison}.

\combinedtable
\subsection{Effect of Prediction Horizon}
The planning horizon significantly impacts the performance of \method (Table \ref{tab:combined}), similar to its effect on conventional planning algorithms, where complex tasks require a sufficiently long planning horizon. With a shorter horizon of 16 timesteps ($\approx 0.5$s), the policy lacks foresight and struggles with forest navigation and complex maneuvers (squatting and jumping), leading to poor success rates. In contrast, extending the horizon to 32 timesteps ($\approx 1$s) dramatically improves performance, with forest navigation success at 96\% and jumping tasks at 71\%. 
\textcolor{black}{Fig. \ref{fig:horizon_trajectories} highlights this comparison visually: a $0.5$s horizon results in circular or stuck behaviors, as the short-sighted policy fails to make progress through the forest, while a $1$s horizon enables successful task completion with diverse and adaptive solution paths that navigate around obstacles effectively. }
\textcolor{black}{However, extending the horizon to $2$s reduces performance, as increased long-term prediction variance causes the policy to overcommit to suboptimal trajectories. As uncertainty decreases near obstacles, the policy recognizes the suboptimality but lacks flexibility to adjust.}

In contrast, a moderate action length improves action quality in agile motions like jumping. While one-step predictions cause jitter, longer predictions face higher future variance. Nevertheless, codiffusing states and actions allows it to leverage the lower state variance to improve the robustness of long action predictions, an advancement over prior works~\cite{xiayou_Diffuseloco, takara_PDP}.
In general, 32 state predictions combined with 16 action predictions yield optimal performance across our evaluation tasks. 

\begin{figure*}[t]
    \centering
    \includegraphics[width=0.9\linewidth]{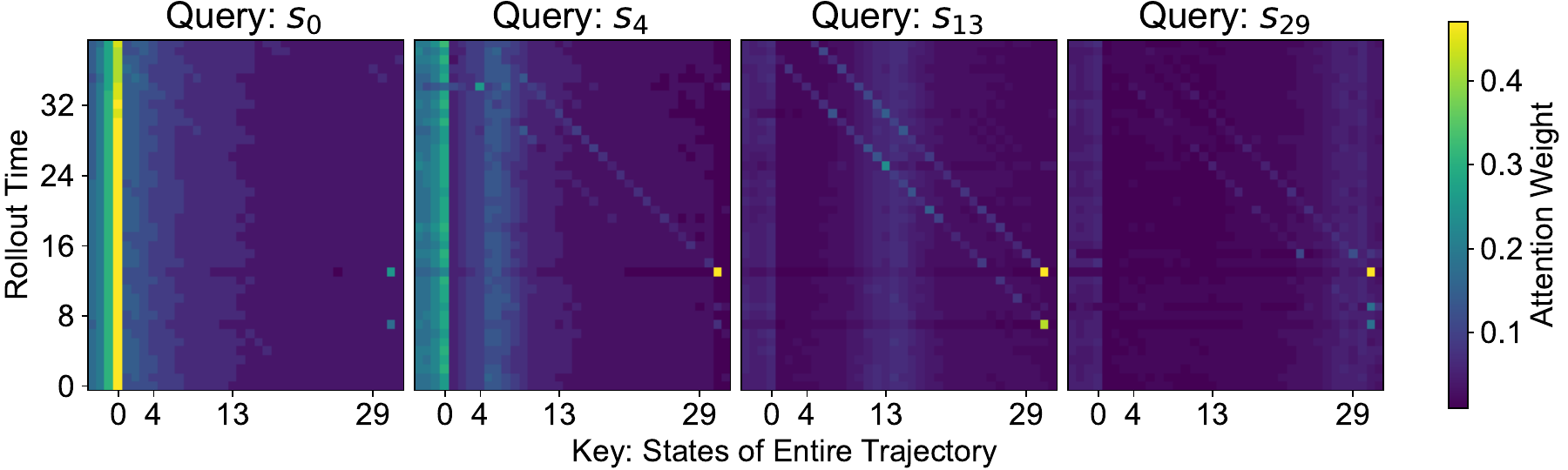}
    \caption{\label{fig:attn} \textbf{Attention Maps for Obstacle Jumping}. Each subplot shows how a state \( s_t \) attends to tokens across the prediction horizon (x axis, $-3$ to $0$ are observations) and over simulation time steps (y axis). First, the model learns localized attention to nearby states over time, indicated by the vertical color band around each state. Diagonal patterns at \( s_4 \), \( s_{13} \), and \( s_{29} \) indicate the model attends to the obstacle when predicted states encounter it with guidance cost and propagates this information backward for planning. Notably, future states also attend to past obstacle events to support planning. 
    In contrast, $s_0$, the current state, consistently attends more strongly to past states than to future predictions, reducing the impact of artifacts from noisy future states. 
    }
    
\end{figure*}

\subsection{Effect of Different Attention Schemes}
The attention mechanism plays a critical role in enhancing policy robustness. To ensure the plan is effectively followed, actions should be influenced by anticipated future conditions. We evaluated different attention schemes to assess their impact on performance.

\textbf{ \textcolor{black}{Full Attention.}} Allowing actions to directly attend to future states (\textcolor{black}{\emph{full attention}}) decreases success rates (Table \ref{tab:combined}). This decrease is due to two factors: (1) future states often include artifacts, and (2) due to the rolling scheme, future states tend to be noisier, making direct attention to future states unreliable.

\textbf{\method's Causal Action Attention.} In contrast, \method adopts \emph{causal attention}, restricting focus to current and past states and actions. Future states still indirectly influence actions by propagating their effects to the current state, \( s_0 \). This approach avoids the noise and artifacts associated with direct future state attention. As shown in Fig. \ref{fig:attn}, \( s_0 \) attends more strongly to past states than to future predictions, prioritizing dependable information and improving generalization. This design enables \method to produce accurate and robust actions across tasks.

\textbf{Diffuser's Local Receptive Field.} Diffuser~\cite{janner_diffuser} employs a limited receptive field, restricting attention to neighboring states and actions. This design prevents effective propagation of future information back to current actions, leading to poor coordination between state predictions and actions. Consequently, Diffuser struggles with task performance, as it cannot align actions with anticipated future conditions.

\subsection{Effect of Different Rolling Schemes}

Shown in Table \ref{tab:combined}, we observe a trade-off in rolling schemes across tasks. In the \emph{jump} task, disabling state rolling causes changes in the motion plan which disrupt the character's momentum, leading to lower jump heights and poor success rates. Conversely, in the \emph{forest} task, an optimal plan for the current timestep may become suboptimal as distant obstacles enter the planning horizon. Enforcing consistency with these suboptimal plans lowers success rates and extends completion time. Despite these trade-offs, our best model performs robustly across scenarios with a single set of parameters. For action rolling, the impact on performance is minimal, enabling a $25\%$ acceleration in the diffusion process without significant loss.




\begin{figure*}[t]
    \centering
    \includegraphics[width=1.0 \linewidth]{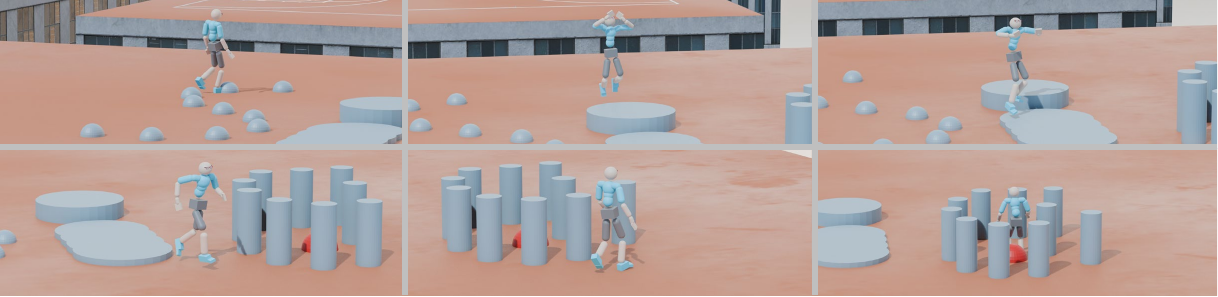}    \caption{\label{fig:downstreamtasks_2} \textbf{A long-horizon sequence of diverse tasks.} A sequence of tasks in a single run, including route path following, platform jumping, and reaching a waypoint behind cylindrical barriers.}
    
\end{figure*}

\section{Discussion and Future Work}
\label{sec:conclusion}
In this work, we introduced \method, a guided diffusion model for physics-based character look-ahead control. By co-diffusing a long horizon of states and actions, our approach overcomes limitations in existing methods, such as limited steerability and physical inconsistencies in generated motions. \method employs inference-time conditioning techniques to leverage intuitive \emph{state-space} guidance for generation in \emph{action space}, enabling flexible reuse for novel downstream tasks without retraining or finetuning. Experimental results highlight its zero-shot capability to address diverse tasks, including static and dynamic obstacle avoidance, dynamic maneuvers, and task-space control using a single model. We showcase a long-horizon run with these diverse tasks in Fig.~\ref{fig:downstreamtasks_2}.

\textcolor{black}{
Despite its effectiveness, our approach has several limitations that open directions for future research. One key challenge is tuning the strength of classifier guidance to ensure task success while staying within the training distribution. For some tasks, particularly those involving jumping or crawling, overly strong guidance sometimes led to unnatural motions or degraded quality. In these cases, increased guidance weight on obstacle clearance was necessary to ensure successful execution. Exploring alternative guidance formulations such as using future returns instead of immediate rewards or incorporating trust region constraints \cite{huang2024constraineddiffusiontrustsampling} could provide better control.}

\textcolor{black}{
Another limitation stems from insufficient data coverage. In tasks like hand-reaching or root path tracking, the model sometimes exhibits erroneous behaviors. For instance, jumping in place when asked to reach overhead. This is because jumping motions are present in the training dataset while overhead-reaching motions are not. While additional data would help mitigate these issues, our experiments were designed to evaluate performance under limited coverage.}

\textcolor{black}{
We also observed reduced motion quality in some results, such as foot jitter. This is likely caused by torque noise added during data augmentation. However, without this noise, diffusion fails due to compounding errors \cite{takara_PDP}. Future work could explore alternative augmentation methods to reduce compounding errors while improving motion smoothness.}

\textcolor{black}{
In addition, the model’s reliance on historical observations biases it toward previously seen motion patterns, which can limit transitions and responsiveness, as noted in \cite{boyuan_diffusionForcing}. For example, some characters in the video are not very responsive to the guided signals in their tasks. We believe historical observations impose a constraint on the character, making the effect of future state predictions propagate slower to affect the next action. Future work could explore reducing this dependency. 
}

\textcolor{black}{
Lastly, extending our approach to human-object interaction remains an open challenge. Physical manipulation and contact-rich behaviors introduce complexities beyond the current framework, presenting an exciting direction for future work. }

\bibliographystyle{ACM-Reference-Format}
\bibliography{bibliography}


\begin{thebibliography}{43}


\ifx \showCODEN    \undefined \def \showCODEN     #1{\unskip}     \fi
\ifx \showDOI      \undefined \def \showDOI       #1{#1}\fi
\ifx \showISBNx    \undefined \def \showISBNx     #1{\unskip}     \fi
\ifx \showISBNxiii \undefined \def \showISBNxiii  #1{\unskip}     \fi
\ifx \showISSN     \undefined \def \showISSN      #1{\unskip}     \fi
\ifx \showLCCN     \undefined \def \showLCCN      #1{\unskip}     \fi
\ifx \shownote     \undefined \def \shownote      #1{#1}          \fi
\ifx \showarticletitle \undefined \def \showarticletitle #1{#1}   \fi
\ifx \showURL      \undefined \def \showURL       {\relax}        \fi
\providecommand\bibfield[2]{#2}
\providecommand\bibinfo[2]{#2}
\providecommand\natexlab[1]{#1}
\providecommand\showeprint[2][]{arXiv:#2}

\bibitem[Ajay et~al\mbox{.}(2023)]%
        {ajay_cond_decision_making}
\bibfield{author}{\bibinfo{person}{Anurag Ajay}, \bibinfo{person}{Yilun Du}, \bibinfo{person}{Abhi Gupta}, \bibinfo{person}{Joshua Tenenbaum}, \bibinfo{person}{Tommi Jaakkola}, {and} \bibinfo{person}{Pulkit Agrawal}.} \bibinfo{year}{2023}\natexlab{}.
\newblock \bibinfo{title}{Is Conditional Generative Modeling all you need for Decision-Making?}
\newblock
\newblock
\urldef\tempurl%
\url{https://openreview.net/forum?id=sP1fo2K9DFG}
\showURL{%
\tempurl}


\bibitem[Carvalho et~al\mbox{.}(2023)]%
        {carvalho2024motionplanningdiffusionlearning}
\bibfield{author}{\bibinfo{person}{Joao Carvalho}, \bibinfo{person}{An~T. Le}, \bibinfo{person}{Mark Baierl}, \bibinfo{person}{Dorothea Koert}, {and} \bibinfo{person}{Jan Peters}.} \bibinfo{year}{2023}\natexlab{}.
\newblock \bibinfo{title}{Motion Planning Diffusion: Learning and Planning of Robot Motions with Diffusion Models}.
\newblock , \bibinfo{numpages}{1916--1923}~pages.
\newblock


\bibitem[Chen et~al\mbox{.}(2024)]%
        {boyuan_diffusionForcing}
\bibfield{author}{\bibinfo{person}{Boyuan Chen}, \bibinfo{person}{Diego~Marti Monso}, \bibinfo{person}{Yilun Du}, \bibinfo{person}{Max Simchowitz}, \bibinfo{person}{Russ Tedrake}, {and} \bibinfo{person}{Vincent Sitzmann}.} \bibinfo{year}{2024}\natexlab{}.
\newblock \bibinfo{title}{Diffusion Forcing: Next-token Prediction Meets Full-Sequence Diffusion}.
\newblock
\newblock
\showeprint[arxiv]{2407.01392}~[cs.LG]
\urldef\tempurl%
\url{https://arxiv.org/abs/2407.01392}
\showURL{%
\tempurl}


\bibitem[Chi et~al\mbox{.}(2023)]%
        {cheng_diffusionPolicy}
\bibfield{author}{\bibinfo{person}{Cheng Chi}, \bibinfo{person}{Zhenjia Xu}, \bibinfo{person}{Siyuan Feng}, \bibinfo{person}{Eric Cousineau}, \bibinfo{person}{Yilun Du}, \bibinfo{person}{Benjamin Burchfiel}, \bibinfo{person}{Russ Tedrake}, {and} \bibinfo{person}{Shuran Song}.} \bibinfo{year}{2023}\natexlab{}.
\newblock \showarticletitle{Diffusion policy: Visuomotor policy learning via action diffusion}.
\newblock \bibinfo{journal}{\emph{The International Journal of Robotics Research}} (\bibinfo{year}{2023}), \bibinfo{pages}{02783649241273668}.
\newblock


\bibitem[Cohan et~al\mbox{.}(2024)]%
        {cohan_flexible_motion_inbetween}
\bibfield{author}{\bibinfo{person}{Setareh Cohan}, \bibinfo{person}{Guy Tevet}, \bibinfo{person}{Daniele Reda}, \bibinfo{person}{Xue~Bin Peng}, {and} \bibinfo{person}{Michiel van~de Panne}.} \bibinfo{year}{2024}\natexlab{}.
\newblock \showarticletitle{Flexible motion in-betweening with diffusion models}. In \bibinfo{booktitle}{\emph{ACM SIGGRAPH 2024 Conference Papers}}. \bibinfo{pages}{1--9}.
\newblock


\bibitem[Dhariwal and Nichol(2024)]%
        {dhariwal_classifierguidance}
\bibfield{author}{\bibinfo{person}{Prafulla Dhariwal} {and} \bibinfo{person}{Alex Nichol}.} \bibinfo{year}{2024}\natexlab{}.
\newblock \showarticletitle{Diffusion models beat GANs on image synthesis}. In \bibinfo{booktitle}{\emph{Proceedings of the 35th International Conference on Neural Information Processing Systems}} \emph{(\bibinfo{series}{NIPS '21})}. \bibinfo{publisher}{Curran Associates Inc.}, \bibinfo{address}{Red Hook, NY, USA}, Article \bibinfo{articleno}{672}, \bibinfo{numpages}{15}~pages.
\newblock
\showISBNx{9781713845393}


\bibitem[Han et~al\mbox{.}(2024)]%
        {Gaoge_ReinDiffuse}
\bibfield{author}{\bibinfo{person}{Gaoge Han}, \bibinfo{person}{Mingjiang Liang}, \bibinfo{person}{Jinglei Tang}, \bibinfo{person}{Yongkang Cheng}, \bibinfo{person}{Wei Liu}, {and} \bibinfo{person}{Shaoli Huang}.} \bibinfo{year}{2024}\natexlab{}.
\newblock \bibinfo{title}{ReinDiffuse: Crafting Physically Plausible Motions with Reinforced Diffusion Model}.
\newblock
\newblock
\showeprint[arxiv]{2410.07296}~[cs.CV]
\urldef\tempurl%
\url{https://arxiv.org/abs/2410.07296}
\showURL{%
\tempurl}


\bibitem[Ho et~al\mbox{.}(2020)]%
        {ho2020denoising}
\bibfield{author}{\bibinfo{person}{Jonathan Ho}, \bibinfo{person}{Ajay Jain}, {and} \bibinfo{person}{Pieter Abbeel}.} \bibinfo{year}{2020}\natexlab{}.
\newblock \showarticletitle{Denoising diffusion probabilistic models}.
\newblock \bibinfo{journal}{\emph{Advances in neural information processing systems}}  \bibinfo{volume}{33} (\bibinfo{year}{2020}), \bibinfo{pages}{6840--6851}.
\newblock


\bibitem[Ho and Salimans(2021)]%
        {jonathan_classifier_free_guidance}
\bibfield{author}{\bibinfo{person}{Jonathan Ho} {and} \bibinfo{person}{Tim Salimans}.} \bibinfo{year}{2021}\natexlab{}.
\newblock \bibinfo{title}{Classifier-Free Diffusion Guidance}.
\newblock
\newblock
\urldef\tempurl%
\url{https://openreview.net/forum?id=qw8AKxfYbI}
\showURL{%
\tempurl}


\bibitem[Huang et~al\mbox{.}(2024b)]%
        {huang2024constraineddiffusiontrustsampling}
\bibfield{author}{\bibinfo{person}{William Huang}, \bibinfo{person}{Yifeng Jiang}, \bibinfo{person}{Tom Van~Wouwe}, {and} \bibinfo{person}{Karen Liu}.} \bibinfo{year}{2024}\natexlab{b}.
\newblock \showarticletitle{Constrained Diffusion with Trust Sampling}. In \bibinfo{booktitle}{\emph{The Thirty-eighth Annual Conference on Neural Information Processing Systems}}.
\newblock


\bibitem[Huang et~al\mbox{.}(2024a)]%
        {xiayou_Diffuseloco}
\bibfield{author}{\bibinfo{person}{Xiaoyu Huang}, \bibinfo{person}{Yufeng Chi}, \bibinfo{person}{Ruofeng Wang}, \bibinfo{person}{Zhongyu Li}, \bibinfo{person}{Xue~Bin Peng}, \bibinfo{person}{Sophia Shao}, \bibinfo{person}{Borivoje Nikolic}, {and} \bibinfo{person}{Koushil Sreenath}.} \bibinfo{year}{2024}\natexlab{a}.
\newblock \showarticletitle{DiffuseLoco: Real-Time Legged Locomotion Control with Diffusion from Offline Datasets}. In \bibinfo{booktitle}{\emph{8th Annual Conference on Robot Learning}}.
\newblock
\urldef\tempurl%
\url{https://openreview.net/forum?id=nVJm2RdPDu}
\showURL{%
\tempurl}


\bibitem[Høeg et~al\mbox{.}(2024)]%
        {høeg2024streamingdiffusionpolicyfast}
\bibfield{author}{\bibinfo{person}{Sigmund~H. Høeg}, \bibinfo{person}{Yilun Du}, {and} \bibinfo{person}{Olav Egeland}.} \bibinfo{year}{2024}\natexlab{}.
\newblock \bibinfo{title}{Streaming Diffusion Policy: Fast Policy Synthesis with Variable Noise Diffusion Models}.
\newblock
\newblock
\showeprint[arxiv]{2406.04806}~[cs.RO]
\urldef\tempurl%
\url{https://arxiv.org/abs/2406.04806}
\showURL{%
\tempurl}


\bibitem[Janner et~al\mbox{.}(2022)]%
        {janner_diffuser}
\bibfield{author}{\bibinfo{person}{Michael Janner}, \bibinfo{person}{Yilun Du}, \bibinfo{person}{Joshua Tenenbaum}, {and} \bibinfo{person}{Sergey Levine}.} \bibinfo{year}{2022}\natexlab{}.
\newblock \showarticletitle{Planning with Diffusion for Flexible Behavior Synthesis}.
\newblock   \bibinfo{volume}{162} (\bibinfo{date}{17--23 Jul} \bibinfo{year}{2022}), \bibinfo{pages}{9902--9915}.
\newblock
\urldef\tempurl%
\url{https://proceedings.mlr.press/v162/janner22a.html}
\showURL{%
\tempurl}


\bibitem[Karunratanakul et~al\mbox{.}(2023)]%
        {Guided-MDM}
\bibfield{author}{\bibinfo{person}{Korrawe Karunratanakul}, \bibinfo{person}{Konpat Preechakul}, \bibinfo{person}{Supasorn Suwajanakorn}, {and} \bibinfo{person}{Siyu Tang}.} \bibinfo{year}{2023}\natexlab{}.
\newblock \showarticletitle{{ Guided Motion Diffusion for Controllable Human Motion Synthesis }}. In \bibinfo{booktitle}{\emph{2023 IEEE/CVF International Conference on Computer Vision (ICCV)}}. \bibinfo{publisher}{IEEE Computer Society}, \bibinfo{address}{Los Alamitos, CA, USA}, \bibinfo{pages}{2151--2162}.
\newblock
\urldef\tempurl%
\url{https://doi.org/10.1109/ICCV51070.2023.00205}
\showDOI{\tempurl}


\bibitem[Li et~al\mbox{.}(2024b)]%
        {Tianyu_AAMDM}
\bibfield{author}{\bibinfo{person}{Tianyu Li}, \bibinfo{person}{Calvin Qiao}, \bibinfo{person}{Guanqiao Ren}, \bibinfo{person}{KangKang Yin}, {and} \bibinfo{person}{Sehoon Ha}.} \bibinfo{year}{2024}\natexlab{b}.
\newblock \bibinfo{title}{AAMDM: Accelerated Auto-regressive Motion Diffusion Model}.
\newblock , \bibinfo{numpages}{1813--1823}~pages.
\newblock


\bibitem[Li et~al\mbox{.}(2024a)]%
        {Zhuo_morph}
\bibfield{author}{\bibinfo{person}{Zhuo Li}, \bibinfo{person}{Mingshuang Luo}, \bibinfo{person}{Ruibing Hou}, \bibinfo{person}{Xin Zhao}, \bibinfo{person}{Hao Liu}, \bibinfo{person}{Hong Chang}, \bibinfo{person}{Zimo Liu}, {and} \bibinfo{person}{Chen Li}.} \bibinfo{year}{2024}\natexlab{a}.
\newblock \showarticletitle{Morph: A Motion-free Physics Optimization Framework for Human Motion Generation}.
\newblock \bibinfo{journal}{\emph{arXiv preprint arXiv:2411.14951}} (\bibinfo{year}{2024}).
\newblock


\bibitem[Luo et~al\mbox{.}(2024)]%
        {luo2024universalhumanoidmotionrepresentations}
\bibfield{author}{\bibinfo{person}{Zhengyi Luo}, \bibinfo{person}{Jinkun Cao}, \bibinfo{person}{Josh Merel}, \bibinfo{person}{Alexander Winkler}, \bibinfo{person}{Jing Huang}, \bibinfo{person}{Kris~M Kitani}, {and} \bibinfo{person}{Weipeng Xu}.} \bibinfo{year}{2024}\natexlab{}.
\newblock \showarticletitle{Universal Humanoid Motion Representations for Physics-Based Control}. In \bibinfo{booktitle}{\emph{The Twelfth International Conference on Learning Representations}}.
\newblock


\bibitem[Luo et~al\mbox{.}(2023)]%
        {PHC}
\bibfield{author}{\bibinfo{person}{Zhengyi Luo}, \bibinfo{person}{Jinkun Cao}, \bibinfo{person}{Alexander Winkler}, \bibinfo{person}{Kris Kitani}, {and} \bibinfo{person}{Weipeng Xu}.} \bibinfo{year}{2023}\natexlab{}.
\newblock \showarticletitle{Perpetual Humanoid Control for Real-time Simulated Avatars}. In \bibinfo{booktitle}{\emph{2023 IEEE/CVF International Conference on Computer Vision (ICCV)}}. \bibinfo{pages}{10861--10870}.
\newblock
\urldef\tempurl%
\url{https://doi.org/10.1109/ICCV51070.2023.01000}
\showDOI{\tempurl}


\bibitem[Mahmood et~al\mbox{.}(2019)]%
        {amass2019}
\bibfield{author}{\bibinfo{person}{Naureen Mahmood}, \bibinfo{person}{Nima Ghorbani}, \bibinfo{person}{Nikolaus~F. Troje}, \bibinfo{person}{Gerard Pons-Moll}, {and} \bibinfo{person}{Michael~J. Black}.} \bibinfo{year}{2019}\natexlab{}.
\newblock \showarticletitle{{AMASS}: Archive of Motion Capture as Surface Shapes}. In \bibinfo{booktitle}{\emph{International Conference on Computer Vision}}. \bibinfo{pages}{5442--5451}.
\newblock


\bibitem[Mothish et~al\mbox{.}(2024)]%
        {birodiff}
\bibfield{author}{\bibinfo{person}{GVS Mothish}, \bibinfo{person}{Manan Tayal}, {and} \bibinfo{person}{Shishir Kolathaya}.} \bibinfo{year}{2024}\natexlab{}.
\newblock \bibinfo{title}{BiRoDiff: Diffusion policies for bipedal robot locomotion on unseen terrains}.
\newblock
\newblock
\showeprint[arxiv]{2407.05424}~[cs.RO]
\urldef\tempurl%
\url{https://arxiv.org/abs/2407.05424}
\showURL{%
\tempurl}


\bibitem[Peng et~al\mbox{.}(2018)]%
        {jason_deepMimic}
\bibfield{author}{\bibinfo{person}{Xue~Bin Peng}, \bibinfo{person}{Pieter Abbeel}, \bibinfo{person}{Sergey Levine}, {and} \bibinfo{person}{Michiel van~de Panne}.} \bibinfo{year}{2018}\natexlab{}.
\newblock \showarticletitle{DeepMimic: Example-guided Deep Reinforcement Learning of Physics-based Character Skills}.
\newblock \bibinfo{journal}{\emph{ACM Trans. Graph.}} \bibinfo{volume}{37}, \bibinfo{number}{4}, Article \bibinfo{articleno}{143} (\bibinfo{date}{July} \bibinfo{year}{2018}), \bibinfo{numpages}{14}~pages.
\newblock
\showISSN{0730-0301}
\urldef\tempurl%
\url{https://doi.org/10.1145/3197517.3201311}
\showDOI{\tempurl}


\bibitem[Peng et~al\mbox{.}(2022)]%
        {jason_ASE}
\bibfield{author}{\bibinfo{person}{Xue~Bin Peng}, \bibinfo{person}{Yunrong Guo}, \bibinfo{person}{Lina Halper}, \bibinfo{person}{Sergey Levine}, {and} \bibinfo{person}{Sanja Fidler}.} \bibinfo{year}{2022}\natexlab{}.
\newblock \showarticletitle{Ase: Large-scale reusable adversarial skill embeddings for physically simulated characters}.
\newblock \bibinfo{journal}{\emph{ACM Transactions On Graphics (TOG)}} \bibinfo{volume}{41}, \bibinfo{number}{4} (\bibinfo{year}{2022}), \bibinfo{pages}{1--17}.
\newblock


\bibitem[Peng et~al\mbox{.}(2021)]%
        {jason_AMP}
\bibfield{author}{\bibinfo{person}{Xue~Bin Peng}, \bibinfo{person}{Ze Ma}, \bibinfo{person}{Pieter Abbeel}, \bibinfo{person}{Sergey Levine}, {and} \bibinfo{person}{Angjoo Kanazawa}.} \bibinfo{year}{2021}\natexlab{}.
\newblock \showarticletitle{AMP: adversarial motion priors for stylized physics-based character control}.
\newblock \bibinfo{journal}{\emph{ACM Trans. Graph.}} \bibinfo{volume}{40}, \bibinfo{number}{4}, Article \bibinfo{articleno}{144} (\bibinfo{date}{July} \bibinfo{year}{2021}), \bibinfo{numpages}{20}~pages.
\newblock
\showISSN{0730-0301}
\urldef\tempurl%
\url{https://doi.org/10.1145/3450626.3459670}
\showDOI{\tempurl}


\bibitem[Radford(2018)]%
        {radford2018improving}
\bibfield{author}{\bibinfo{person}{Alec Radford}.} \bibinfo{year}{2018}\natexlab{}.
\newblock \showarticletitle{Improving language understanding by generative pre-training}.
\newblock  (\bibinfo{year}{2018}).
\newblock


\bibitem[Römer et~al\mbox{.}(2024)]%
        {ralf_dp_with_constraints}
\bibfield{author}{\bibinfo{person}{Ralf Römer}, \bibinfo{person}{Alexander von Rohr}, {and} \bibinfo{person}{Angela~P. Schoellig}.} \bibinfo{year}{2024}\natexlab{}.
\newblock \bibinfo{title}{Diffusion Predictive Control with Constraints}.
\newblock
\newblock
\showeprint[arxiv]{2412.09342}~[cs.RO]
\urldef\tempurl%
\url{https://arxiv.org/abs/2412.09342}
\showURL{%
\tempurl}


\bibitem[Serifi et~al\mbox{.}(2024a)]%
        {agon_robotMDM}
\bibfield{author}{\bibinfo{person}{Agon Serifi}, \bibinfo{person}{Ruben Grandia}, \bibinfo{person}{Espen Knoop}, \bibinfo{person}{Markus Gross}, {and} \bibinfo{person}{Moritz B\"{a}cher}.} \bibinfo{year}{2024}\natexlab{a}.
\newblock \showarticletitle{Robot Motion Diffusion Model: Motion Generation for Robotic Characters}. In \bibinfo{booktitle}{\emph{SIGGRAPH Asia 2024 Conference Papers}} \emph{(\bibinfo{series}{SA '24})}. \bibinfo{publisher}{Association for Computing Machinery}, \bibinfo{address}{New York, NY, USA}, Article \bibinfo{articleno}{50}, \bibinfo{numpages}{9}~pages.
\newblock
\showISBNx{9798400711312}
\urldef\tempurl%
\url{https://doi.org/10.1145/3680528.3687626}
\showDOI{\tempurl}


\bibitem[Serifi et~al\mbox{.}(2024b)]%
        {agon_vmp}
\bibfield{author}{\bibinfo{person}{Agon Serifi}, \bibinfo{person}{Ruben Grandia}, \bibinfo{person}{Espen Knoop}, \bibinfo{person}{Markus Gross}, {and} \bibinfo{person}{Moritz Bächer}.} \bibinfo{year}{2024}\natexlab{b}.
\newblock \showarticletitle{VMP: Versatile Motion Priors for Robustly Tracking Motion on Physical Characters}.
\newblock \bibinfo{journal}{\emph{Computer Graphics Forum}} \bibinfo{volume}{43}, \bibinfo{number}{8} (\bibinfo{year}{2024}), \bibinfo{pages}{e15175}.
\newblock
\urldef\tempurl%
\url{https://doi.org/10.1111/cgf.15175}
\showDOI{\tempurl}


\bibitem[Shi et~al\mbox{.}(2024)]%
        {yi_AMDM}
\bibfield{author}{\bibinfo{person}{Yi Shi}, \bibinfo{person}{Jingbo Wang}, \bibinfo{person}{Xuekun Jiang}, \bibinfo{person}{Bingkun Lin}, \bibinfo{person}{Bo Dai}, {and} \bibinfo{person}{Xue~Bin Peng}.} \bibinfo{year}{2024}\natexlab{}.
\newblock \showarticletitle{Interactive Character Control with Auto-Regressive Motion Diffusion Models}.
\newblock \bibinfo{journal}{\emph{ACM Trans. Graph.}} \bibinfo{volume}{43}, \bibinfo{number}{4}, Article \bibinfo{articleno}{143} (\bibinfo{date}{July} \bibinfo{year}{2024}), \bibinfo{numpages}{14}~pages.
\newblock
\showISSN{0730-0301}
\urldef\tempurl%
\url{https://doi.org/10.1145/3658140}
\showDOI{\tempurl}


\bibitem[Song et~al\mbox{.}(2021)]%
        {song2020score}
\bibfield{author}{\bibinfo{person}{Yang Song}, \bibinfo{person}{Jascha Sohl-Dickstein}, \bibinfo{person}{Diederik~P Kingma}, \bibinfo{person}{Abhishek Kumar}, \bibinfo{person}{Stefano Ermon}, {and} \bibinfo{person}{Ben Poole}.} \bibinfo{year}{2021}\natexlab{}.
\newblock \showarticletitle{Score-Based Generative Modeling through Stochastic Differential Equations}. In \bibinfo{booktitle}{\emph{International Conference on Learning Representations}}.
\newblock


\bibitem[Tevet et~al\mbox{.}(2024)]%
        {guy_closd}
\bibfield{author}{\bibinfo{person}{Guy Tevet}, \bibinfo{person}{Sigal Raab}, \bibinfo{person}{Setareh Cohan}, \bibinfo{person}{Daniele Reda}, \bibinfo{person}{Zhengyi Luo}, \bibinfo{person}{Xue~Bin Peng}, \bibinfo{person}{Amit~H. Bermano}, {and} \bibinfo{person}{Michiel van~de Panne}.} \bibinfo{year}{2024}\natexlab{}.
\newblock \bibinfo{title}{CLoSD: Closing the Loop between Simulation and Diffusion for multi-task character control}.
\newblock
\newblock
\showeprint[arxiv]{2410.03441}~[cs.CV]
\urldef\tempurl%
\url{https://arxiv.org/abs/2410.03441}
\showURL{%
\tempurl}


\bibitem[Tevet et~al\mbox{.}(2023)]%
        {guy_MDM}
\bibfield{author}{\bibinfo{person}{Guy Tevet}, \bibinfo{person}{Sigal Raab}, \bibinfo{person}{Brian Gordon}, \bibinfo{person}{Yoni Shafir}, \bibinfo{person}{Daniel Cohen-or}, {and} \bibinfo{person}{Amit~Haim Bermano}.} \bibinfo{year}{2023}\natexlab{}.
\newblock \showarticletitle{Human Motion Diffusion Model}. In \bibinfo{booktitle}{\emph{The Eleventh International Conference on Learning Representations}}.
\newblock


\bibitem[Truong et~al\mbox{.}(2024)]%
        {takara_PDP}
\bibfield{author}{\bibinfo{person}{Takara~Everest Truong}, \bibinfo{person}{Michael Piseno}, \bibinfo{person}{Zhaoming Xie}, {and} \bibinfo{person}{C.~Karen Liu}.} \bibinfo{year}{2024}\natexlab{}.
\newblock \showarticletitle{PDP: Physics-Based Character Animation via Diffusion Policy}. In \bibinfo{booktitle}{\emph{SIGGRAPH Asia 2024 Conference Papers}} \emph{(\bibinfo{series}{SA ’24})}. \bibinfo{publisher}{ACM}, \bibinfo{pages}{1–10}.
\newblock
\urldef\tempurl%
\url{https://doi.org/10.1145/3680528.3687683}
\showDOI{\tempurl}


\bibitem[Tseng et~al\mbox{.}(2023)]%
        {jonathan_EDGE}
\bibfield{author}{\bibinfo{person}{Jonathan Tseng}, \bibinfo{person}{Rodrigo Castellon}, {and} \bibinfo{person}{Karen Liu}.} \bibinfo{year}{2023}\natexlab{}.
\newblock \showarticletitle{EDGE: Editable Dance Generation From Music}. In \bibinfo{booktitle}{\emph{Proceedings of the IEEE/CVF Conference on Computer Vision and Pattern Recognition (CVPR)}}. \bibinfo{pages}{448--458}.
\newblock


\bibitem[Wang et~al\mbox{.}(2024)]%
        {zhendong_onestepdiffusionpolicyfast}
\bibfield{author}{\bibinfo{person}{Zhendong Wang}, \bibinfo{person}{Zhaoshuo Li}, \bibinfo{person}{Ajay Mandlekar}, \bibinfo{person}{Zhenjia Xu}, \bibinfo{person}{Jiaojiao Fan}, \bibinfo{person}{Yashraj Narang}, \bibinfo{person}{Linxi Fan}, \bibinfo{person}{Yuke Zhu}, \bibinfo{person}{Yogesh Balaji}, \bibinfo{person}{Mingyuan Zhou}, \bibinfo{person}{Ming-Yu Liu}, {and} \bibinfo{person}{Yu Zeng}.} \bibinfo{year}{2024}\natexlab{}.
\newblock \bibinfo{title}{One-Step Diffusion Policy: Fast Visuomotor Policies via Diffusion Distillation}.
\newblock
\newblock
\showeprint[arxiv]{2410.21257}~[cs.RO]
\urldef\tempurl%
\url{https://arxiv.org/abs/2410.21257}
\showURL{%
\tempurl}


\bibitem[Welling and Teh(2011)]%
        {welling2011bayesian}
\bibfield{author}{\bibinfo{person}{Max Welling} {and} \bibinfo{person}{Yee~W Teh}.} \bibinfo{year}{2011}\natexlab{}.
\newblock \showarticletitle{Bayesian learning via stochastic gradient Langevin dynamics}. In \bibinfo{booktitle}{\emph{Proceedings of the 28th international conference on machine learning (ICML-11)}}. \bibinfo{pages}{681--688}.
\newblock


\bibitem[Won et~al\mbox{.}(2022)]%
        {jungdom_characterControlVAE}
\bibfield{author}{\bibinfo{person}{Jungdam Won}, \bibinfo{person}{Deepak Gopinath}, {and} \bibinfo{person}{Jessica Hodgins}.} \bibinfo{year}{2022}\natexlab{}.
\newblock \showarticletitle{Physics-based character controllers using conditional VAEs}.
\newblock \bibinfo{journal}{\emph{ACM Trans. Graph.}} \bibinfo{volume}{41}, \bibinfo{number}{4}, Article \bibinfo{articleno}{96} (\bibinfo{date}{jul} \bibinfo{year}{2022}), \bibinfo{numpages}{12}~pages.
\newblock
\showISSN{0730-0301}
\urldef\tempurl%
\url{https://doi.org/10.1145/3528223.3530067}
\showDOI{\tempurl}


\bibitem[Xie et~al\mbox{.}(2023)]%
        {zhaoming_BoxLoco}
\bibfield{author}{\bibinfo{person}{Zhaoming Xie}, \bibinfo{person}{Jonathan Tseng}, \bibinfo{person}{Sebastian Starke}, \bibinfo{person}{Michiel van~de Panne}, {and} \bibinfo{person}{C.~Karen Liu}.} \bibinfo{year}{2023}\natexlab{}.
\newblock \bibinfo{title}{Hierarchical Planning and Control for Box Loco-Manipulation}.
\newblock , \bibinfo{numpages}{18}~pages.
\newblock


\bibitem[Yao et~al\mbox{.}(2022)]%
        {Yao2022ControlVAE}
\bibfield{author}{\bibinfo{person}{Heyuan Yao}, \bibinfo{person}{Zhenhua Song}, \bibinfo{person}{Baoquan Chen}, {and} \bibinfo{person}{Libin Liu}.} \bibinfo{year}{2022}\natexlab{}.
\newblock \showarticletitle{ControlVAE: Model-Based Learning of Generative Controllers for Physics-Based Characters}.
\newblock \bibinfo{journal}{\emph{ACM Transactions on Graphics}} \bibinfo{volume}{41}, \bibinfo{number}{6} (\bibinfo{date}{Nov.} \bibinfo{year}{2022}), \bibinfo{pages}{1–16}.
\newblock
\showISSN{1557-7368}
\urldef\tempurl%
\url{https://doi.org/10.1145/3550454.3555434}
\showDOI{\tempurl}


\bibitem[Yao et~al\mbox{.}(2024)]%
        {yao2023moconvq}
\bibfield{author}{\bibinfo{person}{Heyuan Yao}, \bibinfo{person}{Zhenhua Song}, \bibinfo{person}{Yuyang Zhou}, \bibinfo{person}{Tenglong Ao}, \bibinfo{person}{Baoquan Chen}, {and} \bibinfo{person}{Libin Liu}.} \bibinfo{year}{2024}\natexlab{}.
\newblock \showarticletitle{MoConVQ: Unified Physics-Based Motion Control via Scalable Discrete Representations}.
\newblock \bibinfo{journal}{\emph{ACM Transactions on Graphics (TOG)}} \bibinfo{volume}{43}, \bibinfo{number}{4} (\bibinfo{year}{2024}), \bibinfo{pages}{1--21}.
\newblock


\bibitem[Zhang et~al\mbox{.}(2023)]%
        {zhang2023learning}
\bibfield{author}{\bibinfo{person}{Yunbo Zhang}, \bibinfo{person}{Alexander Clegg}, \bibinfo{person}{Sehoon Ha}, \bibinfo{person}{Greg Turk}, {and} \bibinfo{person}{Yuting Ye}.} \bibinfo{year}{2023}\natexlab{}.
\newblock \showarticletitle{Learning to Transfer In-Hand Manipulations Using a Greedy Shape Curriculum}. In \bibinfo{booktitle}{\emph{Computer graphics forum}}, Vol.~\bibinfo{volume}{42}. Wiley Online Library, \bibinfo{pages}{25--36}.
\newblock


\bibitem[Zhang et~al\mbox{.}(2024)]%
        {zhang_Tedi}
\bibfield{author}{\bibinfo{person}{Zihan Zhang}, \bibinfo{person}{Richard Liu}, \bibinfo{person}{Rana Hanocka}, {and} \bibinfo{person}{Kfir Aberman}.} \bibinfo{year}{2024}\natexlab{}.
\newblock \showarticletitle{TEDi: Temporally-Entangled Diffusion for Long-Term Motion Synthesis}. In \bibinfo{booktitle}{\emph{ACM SIGGRAPH 2024 Conference Papers}} (Denver, CO, USA) \emph{(\bibinfo{series}{SIGGRAPH '24})}. \bibinfo{publisher}{Association for Computing Machinery}, \bibinfo{address}{New York, NY, USA}, Article \bibinfo{articleno}{68}, \bibinfo{numpages}{11}~pages.
\newblock
\showISBNx{9798400705250}
\urldef\tempurl%
\url{https://doi.org/10.1145/3641519.3657515}
\showDOI{\tempurl}


\bibitem[Zhao et~al\mbox{.}(2023)]%
        {tony_actionChunking}
\bibfield{author}{\bibinfo{person}{Tony~Z. Zhao}, \bibinfo{person}{Vikash Kumar}, \bibinfo{person}{Sergey Levine}, {and} \bibinfo{person}{Chelsea Finn}.} \bibinfo{year}{2023}\natexlab{}.
\newblock \bibinfo{title}{Learning Fine-Grained Bimanual Manipulation with Low-Cost Hardware}.
\newblock
\newblock
\urldef\tempurl%
\url{https://doi.org/10.15607/RSS.2023.XIX.016}
\showDOI{\tempurl}


\bibitem[Zhu et~al\mbox{.}(2023)]%
        {zhu2023neural}
\bibfield{author}{\bibinfo{person}{Qingxu Zhu}, \bibinfo{person}{He Zhang}, \bibinfo{person}{Mengting Lan}, {and} \bibinfo{person}{Lei Han}.} \bibinfo{year}{2023}\natexlab{}.
\newblock \showarticletitle{Neural Categorical Priors for Physics-Based Character Control}.
\newblock \bibinfo{journal}{\emph{ACM Transactions on Graphics (TOG)}} \bibinfo{volume}{42}, \bibinfo{number}{6} (\bibinfo{year}{2023}), \bibinfo{pages}{1--16}.
\newblock


\end{thebibliography}

\clearpage
\appendix


\onecolumn

\section{Task Weights}

Here, we provide the weights for different costs listed in Sec. 4 in computing classifier guidance for each tasks. 
\begin{table*}[htbp]
\centering
\caption{Task Weights Table}
\resizebox{\textwidth}{!}{%
\begin{tabular}{>{\raggedright\arraybackslash}p{4cm}p{3.5cm}p{3cm}ccc p{2.5cm}}
\toprule
\textbf{Task Category} & \textbf{Task} & \textbf{Task Space Ctrl} & \textbf{Static Obs.} & \textbf{Dyn. Obs.} & \textbf{Way-point} & \textbf{Other} \\
\midrule
\multirow{4}{4cm}{Task Space Control} 
  & Root Path Following & 0.2 (root path) & -- & -- & -- & -- \\
  & Reaching & 0.2 (selected body) & -- & -- & -- & -- \\
  & Gamepad Control & 0.1 (velocity) & -- & -- & -- & -- \\
  & Walk-Perturb & 0.1 (forward vel.) & -- & -- & -- & -- \\
\midrule
\multirow{2}{4cm}{Static Obstacle Avoidance}
  & Jump & -- & 1 & -- & 0.06 & -- \\
  & Forest & -- & 1 & -- & 0.175 & -- \\
\midrule
Dynamic Obstacle Avoidance
  & Forest + Dynamic & -- & 1 & 1 & 0.175 & -- \\
\midrule
Motion In-betweening
  & Motion In-betweening & -- & -- & -- & 0.08 & Inpainting \\
\bottomrule
\end{tabular}
}
\label{tab:task-weights}
\end{table*}

\end{document}